\documentclass{aa}

\usepackage{graphicx}
\usepackage{txfonts}
\usepackage[breaklinks,colorlinks,citecolor=blue]{hyperref}
\usepackage{enumitem}
\usepackage{amsmath, amssymb}
\usepackage{natbib}
\usepackage[all]{hypcap}

\begin{document} 

   \title{Toward a consistent use of overshooting parametrizations in 1D stellar evolution codes}

   \titlerunning{Toward a consistent use of overshooting parametrizations}
   \authorrunning{M. Viallet et al.}

   \author{M. Viallet\inst{1}\fnmsep\thanks{\email{mviallet@mpa-garching.mpg.de}}
          \and
          C. Meakin\inst{2,3,4}
          \and
          V. Prat\inst{1}
          \and
          D. Arnett\inst{2}
          }

   \institute{Max-Planck-Institut f\"ur Astrophysik, Karl-Schwarzschild-Str. 1,  85748 Garching, Germany
         \and
         Steward Observatory, 933 N. Cherry Ave., Tucson, AZ 85721, USA
         \and
         New Mexico Consortium, Los Alamos, NM 87544, USA
         \and
         Theoretical Division, Los Alamos National Laboratory, Los Alamos, NM 87545, USA
         }

   \date{Received ; accepted}

   \abstract{Several parametrizations for overshooting in 1D stellar evolution calculations coexist in the literature. These parametrizations are used somewhat arbitrarily in stellar evolution codes, based on what works best for a given problem or even for the historical reasons related to the development of each code. We point out that these different parametrizations correspond to different physical regimes of overshooting, depending on whether the effects of radiation are dominant, marginal, or negligible. Our analysis is based on previously published theoretical results, as well as on multidimensional hydrodynamical simulations of stellar convection where the interaction between the convective region and a stably stratified region is observed. Although the underlying hydrodynamical processes are the same, the outcome of the overshooting process is profoundly affected by radiative effects. Using a simple picture of the scales involved in the overshooting process, we show how three regimes are obtained, depending on the importance of radiative effects. These three regimes correspond to the different behaviors observed in hydrodynamical simulations so far and to the three types of parametrizations used in 1D codes. We suggest that the existing parametrizations for overshooting should coexist in 1D stellar evolution codes and should be applied consistently at convective boundaries depending on the local physical conditions.}
   

   \maketitle
%

\section{Introduction}
\label{sect:introduction}

For more than 50 years, convection has been described in 1D stellar evolution codes by the mixing-length theory \citep[MLT, cf.][]{bohm-vitense_uber_1958}. It is well known that one of the shortcomings of MLT is its inability, by construction, to describe the boundary layer between the convective region and a neighboring stably stratified region. In the MLT picture, the flow simply stops at the boundary. This contradicts basic physics, because inertia allows the flow to penetrate the stably stratified region, inducing mixing beyond the boundary of the convective region \citep[see, e.g.,][]{arnett_321D_2015}.

Such an extra mixing at convective boundaries is routinely added in stellar evolution calculations, since it is required to reproduce well-established observational features across the Hertzsprung-Russel diagram \citep[e.g.,][]{maeder_stellar_1975,matraka_overshooting_1982,schroder_critical_1997,herwig_review_2005,pace_lithium_2012,montalban_rg_2013}.  Currently, this extra mixing is parametrized in very crude ways in stellar evolution codes, and it relies on free parameters that need to be calibrated. This situation is very similar to the one of convection itself, with MLT relying on the mixing length parameter. In fact, overshooting is one aspect of convection, and a better description of overshooting should eventually rely on a better description of turbulent convection. A lot of effort has been devoted to the development of a non-local theory of convection, which, in principle, should allow for a consistent description of overshooting \citep{gough_mixing-length_1977,stellingwerf_convection_1982-1,xiong_evolution_1986,kuhfuss_model_1986,canuto_stellar_1991,canuto_turbulent_1992,gehmeyr_new_1992,wuchterl_simple_1998,deng_anisotropic_2006}. Usually, this is done by looking for a working closure in the framework of Reynolds-Averaged Navier-Stokes (RANS) equations (see e.g. \cite{canuto_compressible_1997,xiong_nonlocal_1997,canuto_stellar_1998,deng_anisotropic_2006,canuto_stellar_2011-3} and references therein). 

Although RANS is the appropriate mathematical framework for improving the description of hydrodynamical processes in stellar evolution, one should abandon the idea of finding the ``ultimate'' parametrization that would describe overshooting correctly in any physical situations. Although the word ``overshooting'' is widely used to describe extra mixing at convective boundaries (and we adopt this terminology as well), it is misleading because the mixing cannot be seen simply as resulting from a unique, basic, physical process. Mixing at convective boundaries likely results from the interplay of several physical processes (shear instabilities, g-modes, etc.), the usual picture of plume penetration being only one aspect of the problem. A better description of convective boundary mixing requires first of all a better understanding of the nature and efficiency of the mixing processes that are taking place in the boundary layer and how they are affected by the local physical conditions. This work takes a first step in this direction by looking specifically at the effect of thermal diffusion. 

In Sect. \ref{sect:params}, we describe the three types of parametrization for overshooting that are used in stellar evolution calculations. Currently, these parametrizations are used somewhat arbitrarily.  In Sect. \ref{sect:nonadiabatic}, we develop a physical picture for the effects of radiation on the overshooting process that leads us to define three distinctive regimes of overshooting. Hydrodynamical simulations support our picture and are discussed. Furthermore, we show that the different parametrizations that are used currently correspond to these different thermal regimes of overshooting. We conclude in Sect. \ref{sect:conclusion} that the description of overshooting in stellar evolution codes could be already improved by using existing parametrizations in a more physically consistent way. In case such a better practice leads to theoretical predictions contradicting observations for some specific cases, this would point to shortcomings in the current descriptions and provide valuable information for improving them.






\section{Existing parametrizations of overshooting}
\label{sect:params}


\subsection{``Chemical mixing''}
\label{sect:diffusive}


In this approach, it is assumed that overshooting does not affect the thermal structure. In terms of the logarithmic temperature gradients, one has $\nabla = \nabla_\mathrm{rad}$ in the overshooting layer. Regarding the mixing of chemical elements, two approaches are possible. In the simplest approach, the composition is assumed to be mixed instantaneously over a distance $l_\mathrm{ov}$ beyond the limit of the convective region, with $l_\mathrm{ov}$ typically parametrized in terms of the pressure scale height. This is the approach taken in BaSTI \citep{basti_2004}, TGEC \citep{TGEC_2012}, and DSEP \citep{DSEP_2008}. Another possibility is to treat this mixing as a diffusive process with a prescribed diffusion coefficient $D_\mathrm{ov}$. Originally, this approach was suggested by \cite{freytag_overshooting_1996} based on radiative hydrodynamical simulations of near-surface convection. Following \cite{freytag_overshooting_1996}, $D_\mathrm{ov}$ is usually parametrized as

\begin{equation}
  D_\mathrm{ov} = D_0 \exp{\left( - 2 \frac{r - r_0}{f H_p} \right)},
\end{equation}

\noindent with $r_0$ the location of the convective boundary, which is typically given by the Schwarzschild or Ledoux criteria, $D_0$, which is a reference diffusivity (typically the mixing-length diffusivity $D_\mathrm{MLT}$ evaluated close to the boundary\footnote{By construction $D_\mathrm{MLT}$ is zero at $r=r_0$, so that in practice it is evaluated ``close'' to the boundary. The definition of $D_0$ likely varies between different stellar evolution codes.}), and $f$ a free parameter. This ``diffusive mixing'' approach is the one taken in MESA \citep{mesa_2011} and GARSTEC \citep{GARSTEC_2008}.


\subsection{``Penetrative convection''}
\label{sect:penetrative}

\cite{zahn_convective_1991} showed that when the convective boundary is located deep inside the star, the overshooting material is able to change the entropy stratification and induce a nearly adiabatic (yet subadiabatic) region \citep[see also][]{schmitt_overshoot_1984}. \cite{zahn_convective_1991} coined this ``penetrative convection''. In stellar evolution codes, this is implemented by artificially increasing the size of the adiabatic region on a distance $l_\mathrm{ov}$, which is typically parametrized in terms of the pressure scale height\footnote{Some parametrizations use a maximum limit to prevent problems due to the increasing value of $H_p$ toward the center.}. The major difference from chemical mixing is that the thermal structure of the star is modified: $\nabla = \nabla_\mathrm{ad}$ in the overshooting region. In addition, the chemical composition is assumed to be mixed instantaneously in this region. This treatment of overshooting is the one implemented in GENEC \citep{GENEC_2008}.

\subsection{``Turbulent entrainment''}
\label{sect:entrainment}

Turbulent entrainment is a process that is well-known in geophysics, because it is observed both in the atmosphere and in the oceans. Entrainment is the physical process by which turbulent eddies entrain mass at the convective boundary, inducing a steady growth in the size of the convective region as long as energy is supplied to the system. Turbulent entrainment is observed in the simulations of the oxygen-burning shell presented in \cite{meakin_turbulent_2007}, showing that the relevance of this process for stellar evolution has been overlooked. To our knowledge, only \cite{staritsin_core_2013,staritsin_2014} has published stellar evolution models that include turbulent entrainment at convective boundaries. His approach follows ``bulk'' entrainment models in which the boundary layer is collapsed to a discontinuity separating the well-mixed region from the stably stratified region \citep[see, e.g.,][]{fernando_mixing_1991}. The velocity $V_e$ at which this boundary moves as a result of entrainment can be parametrized as

\begin{equation}
  \label{eq:entrainment_law}
  \frac{V_e}{V_t} = A \mathrm{Ri}_B^{-n},
\end{equation}

\noindent where $V_t$ is the typical turbulent velocity at the boundary, $A$ and $n$ are parameters that characterize the entrainment (see below), and $\mathrm{Ri}_B$ is the so-called bulk Richardson number, which characterizes the ``stiffness'' of the boundary. It is defined as

\begin{equation}
  \mathrm{Ri}_B = \frac{l \Delta b}{V_t^2},
\end{equation}

\noindent where $l$ is the typical size of the turbulent eddies doing the entrainment, and

\begin{equation}
\Delta b = \int N^2 dr
\end{equation}

\noindent is the buoyancy jump across the interface ($N$ is the Brunt-V\"ais\"al\"a frequency). The spatial integration is performed on a region that contains the convective boundary. Once $V_e$ is determined, the size by which the convective region is extended because of entrainment during one time step is given by $d = V_e  \Delta t$.

The entrainment law (\ref{eq:entrainment_law}) is well-established in fluid dynamics \citep[see, e.g.,][]{fernando_mixing_1991}. The values of $A$ and $n$ were measured in different experimental setups and geophysical environments. \cite{meakin_turbulent_2007} found that their simulations correspond to $n \sim 1.05$ and $A \sim 0.027$, which are the values used in \cite{staritsin_core_2013,staritsin_2014}.

\section{Impact of radiation on overshooting}
\label{sect:nonadiabatic}

\subsection{Measure of radiative effects in stellar hydrodynamics}

A characteristic of stellar hydrodynamics is that, at the typical temperature and density of stellar plasma, photons are very efficient in transporting heat. A measure of the effect of radiation on the flow is given by the P\'eclet number Pe, defined as

\begin{equation}
\mathrm{Pe} = \frac{\mathrm{time\ scale\ for\ radiative\ transport\ of\ heat}}{\mathrm{time\ scale\ for\ advective\ transport\ of\ heat}}.
\end{equation}

\noindent When Pe $\gg 1$, radiation has a negligible impact and the flow can be considered as evolving adiabatically. When Pe $\ll 1$, radiation dominates. In the optically thick interior, heat transport by radiation can be described as a diffusion process, and the corresponding thermal diffusivity $\chi$ (units: cm$^2$/s) is given by

\begin{equation}
\chi = \frac{16 \sigma T^3}{3 \kappa \rho^2 c_p},
\end{equation}

\noindent where $T$ is the temperature, $\rho$ the density, $\kappa$ the Rossland opacity, $c_p$ the heat capacity at constant pressure \citep[see, e.g.,][]{kippenhahn_book_2012}. In this case, the P\'eclet number can be defined as

\begin{equation}
\label{eq:peclet}
\mathrm{Pe} = \frac{u l}{\chi},
\end{equation}

\noindent where $u$ (resp. $l$) is a typical velocity (resp. length) scale of the flow. The Prandtl number, defined as the ratio of  viscosity to thermal diffusion Pr=$\nu/\chi$, takes very low values in stellar plasma, typically Pr=$10^{-9}-10^{-6}$. The situation is very different in geophysical flows, where Prandtl numbers are of order unity. In fact, many insights from geophysical studies can be generalized to the stellar case by introducing the effects of thermal diffusion.


\begin{figure}[t] 
   \centering
   \includegraphics[width=0.8\linewidth, trim= 0 0 0 0]{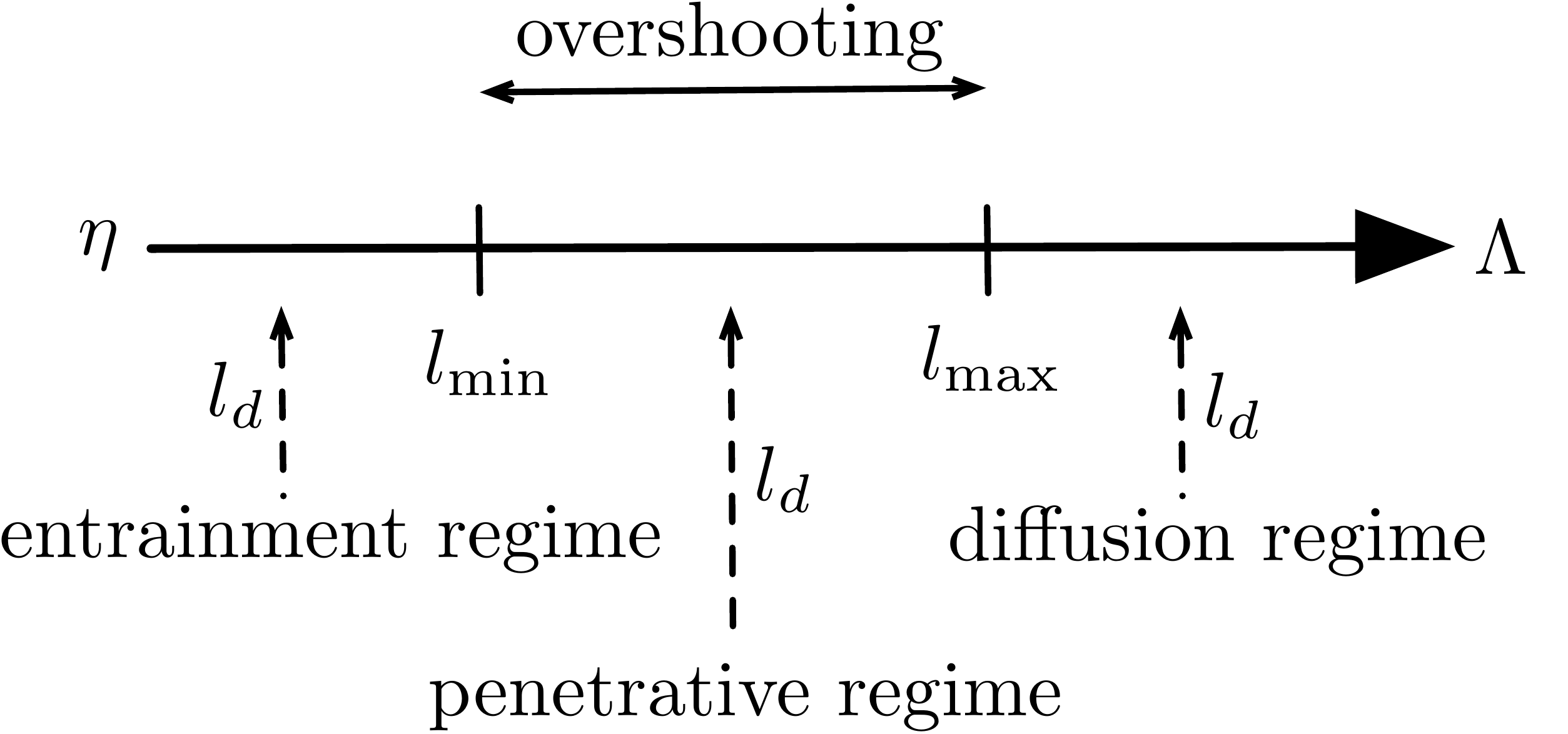}
   \caption{Sketch of the different scales of eddies present in the flow. $\eta$ and $\Lambda$ are the Kolmogorov and integral scales, respectively. The scales contributing to overshooting are those between $l_\mathrm{min}$ and $l_\mathrm{max}$. Depending on the location of the transition scale $l_d$ where $\mathrm{Pe}\sim1$, one can define three different regimes of overshooting (see text).}
   \label{fig:spectrum}
\end{figure}

\subsection{Scales involved in the overshooting process}

Stellar convection is characterized by very high values of the Reynolds number, implying that the flow is highly turbulent: motion extends over a wide range of scales \citep[see, e.g.,][]{arnett_aipa_2014}. An important characteristic of turbulent convection is that transport properties in the bulk of the convective region are dominated by large-scale, coherent structures usually described as ``convective plumes'' \citep[e.g.,][]{stein_topology_1989,cattaneo_turbulent_1991,porter_three-dimensional_2000,murphy_2011,viallet_convection_2013}.

As convective plumes approach the boundary of the convective region, they are deflected sideways. At the same time, inertia allows a plume to penetrate the stably stratified region, where it undergoes ``buoyancy braking'', which is the process in which kinetic energy is converted into potential energy due to the work done against gravity in the stably stratified region. Although buoyancy braking reduces the kinetic energy of the overshooting material, the flow comes to a rest mostly due to the turbulent dissipation that arises as it breaks apart and forms a turbulent cascade through pressure-strain effects resulting from the interaction with the stable region. We coin this process ``buoyancy br\emph{ea}king''.


If we adopt the picture of a turbulent flow extending from the integral scale $\Lambda$ down to the Kolmogorov dissipation scale $\eta$, it is sound to assume that overshooting is effectively achieved by a certain subrange of scales comprised between $l_\mathrm{min}$ and $l_\mathrm{max}$ (see Fig. \ref{fig:spectrum}). Here, $l_\mathrm{max}$ is the typical length scale of plumes, while $l_\mathrm{min}$ is the length scale of the smallest eddies that are able to contribute significantly to mixing at the boundary. We can expect that these scales are distributed spatially: the smallest scales are found deeper in the overshooting region. This physical picture of the scales involved in overshooting is the basis for the interpretation of radiative effects outlined in the next section.

\subsection{Regimes of overshooting depending on the importance of radiative effects}

A turbulent flow is characterized by a wide range of eddies with velocity scale $u$ and length scale $l$. Although the proper definition of a ``turbulent P\'eclet number'' is not trivial, we can consider Eq. (\ref{eq:peclet}) as a qualitative definition of a scale-dependent P\'eclet number. A critical scale is the scale $l_d$ that corresponds to $\mathrm{Pe}\sim1$. Below this scale, eddies are strongly affected by thermal diffusion, above this scale, eddies evolve adiabatically. The assumption that the transition occurs exactly at $\mathrm{Pe}=1$ is a simplification, but it is sufficient for the picture presented in this work. Given the range of scales $[l_\mathrm{min}, l_\mathrm{max}]$ that contribute to overshooting, we define three different regimes of overshooting depending on the location of the transition $\mathrm{Pe}\sim1$, as summarized in Fig. \ref{fig:spectrum} and detailed below.

\subsubsection{Case 1: $l_\mathrm{max} < l_d$ - Diffusion-dominated regime}

In this case, all the scales participating in the overshooting process have $\mathrm{Pe} \lesssim  1$ and are strongly affected by thermal diffusion. Turbulent eddies are only able to mix composition, without affecting the entropy structure  significantly
\citep[see ][]{zahn_convective_1991}. This is the spirit of the ``chemical mixing'' parametrization presented in Sect. \ref{sect:diffusive}. This parametrization was first proposed in \cite{freytag_overshooting_1996}. \cite{freytag_phd_1995} presents radial profiles of the P\'eclet number for the A-type star and white dwarf models published in \cite{freytag_overshooting_1996} (see their Figs. 11 \& 12). They show that $\mathrm{Pe} \lesssim 10$ across the convective and overshooting region, which is consistent with our picture.



\subsubsection{Case 2: $l_\mathrm{min} < l_d < l_\mathrm{max}$ - Penetrative regime}

\begin{figure}[t] 
   \centering
   \includegraphics[width=0.8\linewidth, trim = 20 20 20 20]{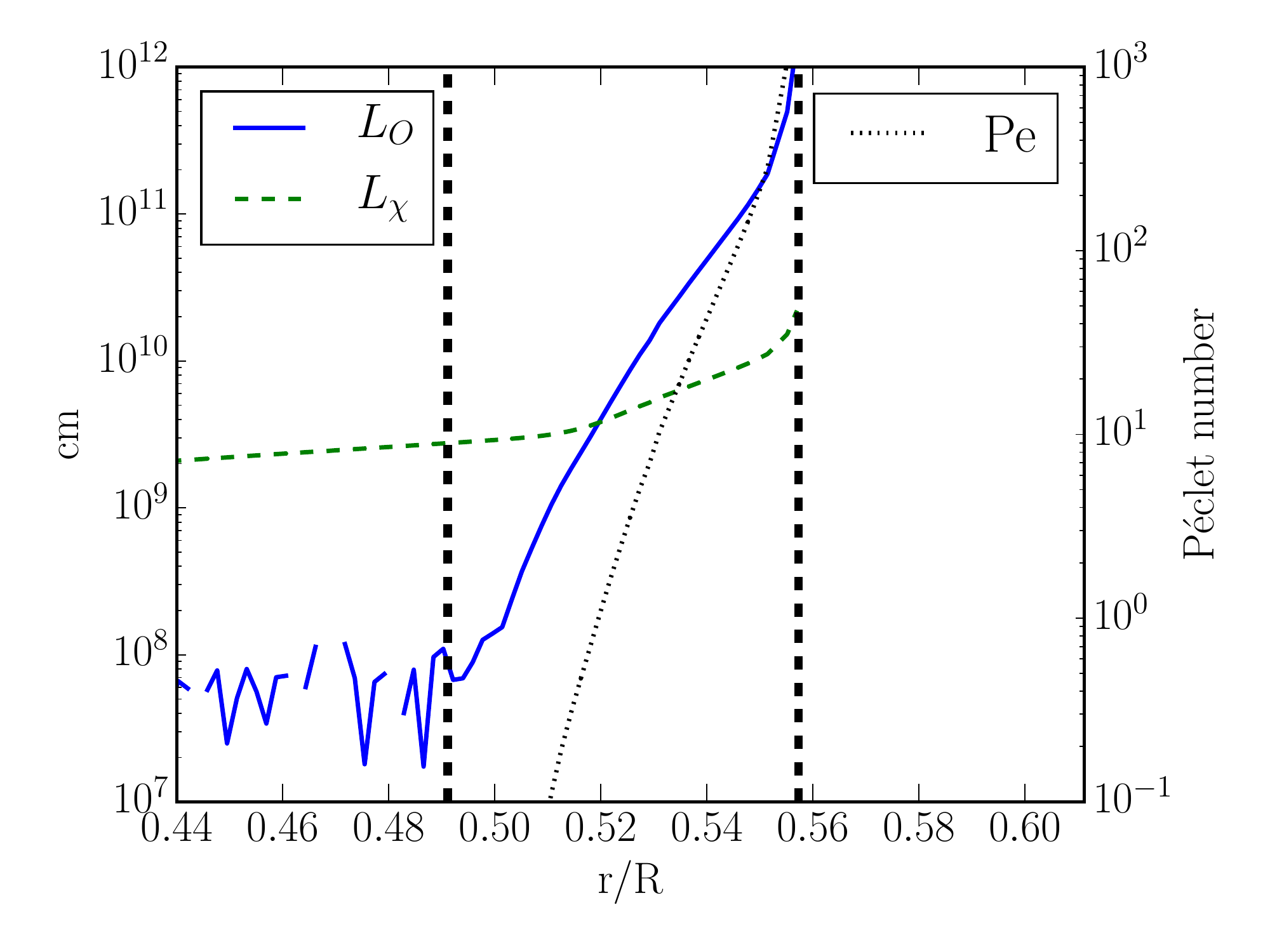} 
   \caption{Radial profiles of the length scales $L_O$, $L_\chi$, and the P\'eclet number in the boundary layer of the red giant model of \cite{viallet_convection_2013}. The thick dashed lines denote the extent of the boundary layer, defined as the region of buoyancy braking. The $x$-axis is normalized by the radius of the star $R = 4.1\times 10^{12}$ cm.}
   \label{fig:rg_lengthscales}
\end{figure}

In this case, there is a transition within the boundary layer: scales that have $\mathrm{Pe}\gtrsim1$ mix entropy and composition, whereas scales that have $\mathrm{Pe}\lesssim1$ are only able to mix composition. This structures the boundary layer in two parts: a nearly adiabatic, well-mixed sublayer, and a ``diffusion-dominated'' sublayer where only the composition is mixed \citep[see][]{zahn_convective_1991}. This picture is supported by the structure of the boundary layer at the bottom of the convective envelope of the red giant model described in \cite{viallet_convection_2013}. Studying the mean-field budget equation for internal energy, the authors find that the boundary layer is in a thermal balance in which the heat advected by entropy-rich material into the layer is counterbalanced by the cooling by radiative diffusion. This results locally in a ``super-stellar" radiative flux ($L_\mathrm{rad} > L_\star$).  

We now show that the boundary layer is indeed characterized by a transition in the turbulent P\'eclet number, consistent with the picture outlined above. We first introduce the so-called Ozmidov scale:

\begin{equation}
L_O = \sqrt{\frac{\epsilon}{N^3}},
\end{equation}

\noindent where $\epsilon$ is the turbulent dissipation rate of kinetic energy, $N$ the Brunt-V\"ais\"al\"a frequency, and $L_O$ is the length scale of the smallest eddies that are affected by buoyancy \citep[e.g.,][]{ozmidov_1965, smyth_length_2000}. We then introduce the length scale resulting from equating the thermal diffusion time scale to the buoyancy time scale:

\begin{equation}
\frac{L_\chi^2}{\chi} = N^{-1} \Leftrightarrow L_\chi = \sqrt{\frac{\chi}{N}},\end{equation}

\noindent where $L_\chi$ can be interpreted as the scale below which thermal diffusion ``erases'' the effect of buoyancy. Inspired by the literature on turbulent mixing \citep{ozmidov_1965,osborn_estimates_1980,brethouwer_numerical_2009}, we take $\epsilon/N^2$ as a proxy for the turbulent transport of heat. One therefore has

\begin{equation}
\label{eq:turbulent_peclet}
\left ( \frac{L_O}{L_\chi} \right)^2 = \frac{\epsilon}{\chi N^2} \sim \mathrm{Pe}.
\end{equation}

\noindent Figure \ref{fig:rg_lengthscales} shows the profiles of $L_O$, $L_\chi$, and Pe in the boundary layer of the red giant model of \cite{viallet_convection_2013}. There is indeed a transition in the P\'eclet number in the boundary layer from $\sim10^3$ to $\sim 10^{-1}$, where the length scales $L_O$ and $L_\chi$ cross each other.

The existence of such a transition is key because it allows a balance to be reached: overshooting extends the size of the adiabatic region until radiative effects inhibit the process\footnote{This local thermal balance is reached quite rapidly, namely on a few dynamical time scales. Naturally, as the stellar structure evolves globally on a Kelvin-Helmholtz (thermal) time scale, the size of the overshooting region will evolve.}. We are currently not able to predict the depth at which this boundary layer develops, since it would require a model for the turbulent dissipation of plumes. Theoretical predictions for the extent of the penetration region exist \citep[e.g.,][]{schmitt_overshoot_1984,zahn_convective_1991,hurlburt_penetration_1994-1,rempel_overshoot_2004}, but they are based on the rather laminar picture of plumes decelerating due to buoyancy braking alone. For the moment, one can rely on the parametrization for ``penetrative convection'' described in Sect. \ref{sect:penetrative}. This parametrization does not include the diffusion-dominated layer in which composition is mixed. \cite{zahn_convective_1991} argues that this layer is very thin \citep[see, however,][]{rempel_overshoot_2004}. In the red-giant model of \cite{viallet_convection_2013}, it occupies roughly one third of the boundary layer. This could result from the limited resolution inherent to numerical simulations of stellar interiors and requires further investigation.
  
Finally, it should be mentioned that \cite{brummell_penetration_2002-4} do not find any evidence of penetrative convection in their 3D simulations. This could be due to their somewhat artificial setup, in which the profile of thermal conductivity is fixed and/or to the fact that their direct numerical simulations probe a different dynamical regime given the values of the dimensionless numbers (Prandtl, Reynolds, Rayleigh numbers) that they could afford in their investigation \citep{rempel_overshoot_2004}.

\subsubsection{Case 3: $l_d < l_\mathrm{min}$ - Entrainment regime}

In this case, all relevant scales have $\mathrm{Pe} \gg 1$, and the overshooting process can be considered to be completely adiabatic. The process of mass entrainment results in a growth of the well-mixed (entropy+composition) region. The key difference with the previous case is that there is nothing that can counterbalance the process: entrainment proceeds as long as energy is injected into the system.

This regime is observed by \cite{meakin_turbulent_2007} in their calculations of the convection in the oxygen-burning shell of a massive star. The reason is that late phases of stellar evolution are driven by neutrino cooling, which acts on a time scale shorter than photon cooling \citep{arnett_supernovae_1996}. As a result, radiative diffusion becomes negligible and does not affect overshooting. Although their boundary layers have finite thickness with a non-trivial internal structure, the authors have shown that their data is described well by the parametrization of ``bulk'' entrainment presented in Sect.~\ref{sect:entrainment}. 

Turbulent entrainment is also observed in hydrodynamical simulations of the He-flash by \cite{mocak_2009,mocak_2010,mocak_2011}. This specific phase of stellar evolution is strongly out of thermal equilibrium owing to the ignition of helium in a degenerate environment. The large release of nuclear energy drives convection on a time scale that is much shorter than the radiative time scale. Radiative diffusion becomes negligible and turbulent entrainment is obtained, as in the previous case.

\section{Conclusion}
\label{sect:conclusion}

Based on previous theoretical and numerical work, this paper presents a simple physical picture of the effects of radiation on overshooting. Other non-adiabatic effects exist in stars (nuclear burning, neutrino losses), but from current simulations we do not see any evidence of a direct role for these effects on overshooting. We show that the three existing parametrizations for overshooting correspond to different regimes, depending on the importance of radiative effects. As a result, we suggest that these three different parametrizations should coexist in 1D stellar evolution codes and be applied consistently depending on the local physical conditions. Ideally, the selection of the adequate parametrization should be based on the value of the P\'eclet number in the boundary layer. However, one faces two difficulties:

\begin{enumerate}
\item How do we estimate the P\'eclet number? According to our work, the quantity that allows distinguishing between the different thermal regimes is given by Eq. (\ref{eq:turbulent_peclet}). However, this quantity cannot be computed in the framework of MLT, which has, by construction, $\epsilon=0$ (no flow) when $N^2>0$ . The P\'eclet number within the convective region can be estimated easily with the formula
\begin{equation}
\mathrm{Pe} = \frac{3 D_\mathrm{MLT}}{\chi},
\end{equation} 

\noindent where $D_\mathrm{MLT} = \frac{1}{3} u_\mathrm{MLT} l_\mathrm{MLT}$ is the usual diffusion coefficient computed from MLT. In a first approach, this quantity could be used to extrapolate the value of the P\'eclet number in the boundary layer.
\item Given a definition of the P\'eclet number at the boundary, what are the range of values that define each regime? The simple picture developed in this work allows us to distinguish between three thermal regimes of overshooting, depending on whether the effect of radiation are dominant, marginal, or negligible. However, it does not make any prediction about the transition between these regimes, whether in terms of the values of the P\'eclet number at with such transitions occur or in terms of how sharp such transitions are.
\end{enumerate}

Therefore, we are not yet able to give a practical criterion that would allow automatic selection of the appropriate description during a stellar evolution calculation. However, as a rule of thumb, we suggest

\begin{enumerate}
\item applying ``chemical mixing'' near the surface of stars, where radiative effects become important (inefficient convection);
\item applying ``penetrative convection'' in the deep interior of stars (efficient convection) for the phases which are photon-cooled (core-overshooting on the main-sequence, convective envelope undershooting, etc.);
\item applying ``turbulent entrainment'' in the deep interior when the evolution is driven by neutrino losses (late stages of stellar evolution) or in phases that are in strong thermal imbalance (e.g., He-flash in low-mass stars).
\end{enumerate}

It should be stressed that we do not claim that the current parametrizations describe each regime accurately. In particular, the calibration of free parameters is still required. However, a good starting point is to use the relevant parametrization for each regime. For instance, it is very likely incorrect to apply turbulent entrainment to describe convective cores on the main sequence as done in \cite{staritsin_core_2013}, since in this phase the time scales are long enough for radiative diffusion to affect overshooting. Likewise, considering our current understanding, it is inconsistent to use the ``chemical mixing'' description deep inside the stellar interior; this may contribute to the errors found by \cite{schindler_2015}. 

Failure to reproduce observational trends should then point to flaws in the parametrizations of these different regimes, such as those due to missing physics. For instance, the effects of composition are currently taken into account very poorly, mainly because of the (arbitrary) choice of using the Schwarzschild or the Ledoux criteria and the ad hoc use of thermohaline mixing. In fact, like radiation, composition probably affects the regime of overshooting and could drastically change the simple picture of the three regimes defined in this work. Likewise, rotation and magnetic field could also affect the nature and the efficiency of mixing at convective boundaries. 

Here as well, insight from multidimensional simulations should help us to understand which are the relevant parameters that characterize these processes and how the parameter space is split into different regimes of overshooting. This work was a first step toward this goal, and it was focused on thermal diffusion. Applying this strategy to other physical processes  should bring us closer to a physically consistent description of convective boundary mixing in 1D stellar evolution codes.




\begin{acknowledgements}
This work is supported by the European Research Council through grant ERC-AdG No. 341157-COCO2CASA. This work used the Extreme Science and Engineering Discovery Environment
(XSEDE), which is supported by National Science Foundation grant number
OCI-1053575. CM and WDA acknowledge support from NSF grant 1107445 at the
University of Arizona. MV and VP thank Francois Ligni\`eres for enlightening discussions. We thank Marcelo Miller Bertolami, Achim Weiss, and Simon Campbell for useful comments on an earlier draft, and Bernd Freytag for pointing out the relevant figures in his thesis.
\end{acknowledgements}

\bibliographystyle{aa}
\bibliography{references}

\end{document}